\documentclass[11pt]{article}

\usepackage[final]{acl}

\usepackage{times}
\usepackage{latexsym}
\usepackage{booktabs}
\usepackage{multirow}
\usepackage{makecell}
\usepackage{tabularx}
\usepackage{colortbl} 
\usepackage{booktabs}
\usepackage{multirow}
\usepackage{array}
\usepackage[table]{xcolor}
\usepackage{colortbl}
\PassOptionsToPackage{hidelinks}{hyperref}

\newcommand{\up}[1]{\textcolor{teal}{$+#1$}}
\newcommand{\dn}[1]{\textcolor{red}{$-#1$}}

\newcolumntype{L}[1]{>{\raggedright\arraybackslash}p{#1}}
\newcolumntype{C}[1]{>{\centering\arraybackslash}p{#1}}
\newcolumntype{Y}{>{\raggedright\arraybackslash}X}

\newcommand{\upb}[1]{\textcolor{teal}{$\mathbf{+#1}$}} % bold = largest gain per model

\usepackage[T1]{fontenc}
\usepackage[utf8]{inputenc}

\usepackage{microtype}

\usepackage{inconsolata}

\usepackage{graphicx}

\newcommand{\affilsize}{\fontsize{11}{11}\selectfont}

\title{Fine-Tuning Large Language Models for Codebook-Guided Coding of Students' Mathematics Metaphor Responses \thanks{This paper has been officially accepted for presentation at the
NCME Artificial Intelligence in Measurement and Education Conference
(AIME-Con) 2026, held October 5--7, 2026, in Pittsburgh, Pennsylvania.}}

\author{
Liang Zhang \\
{\affilsize Department of Learning Health Sciences} \\
{\affilsize University of Michigan--Ann Arbor} \\
{\affilsize \texttt{zhlian@umich.edu}}
\And
Stephen Hwang \\
{\affilsize Department of Mathematical Sciences} \\
{\affilsize University of Delaware} \\
{\affilsize \texttt{hwangste@udel.edu}}
\AND
Yue Ma \\
{\affilsize School of Education} \\
{\affilsize University of Delaware} \\
{\affilsize \texttt{yuemajoy@udel.edu}}
\And
Jinfa Cai \\
{\affilsize Department of Teaching, Learning and Culture} \\
{\affilsize Texas A\&M University} \\
{\affilsize \texttt{jfcai@tamu.edu}}
}
\begin{document}
\maketitle
\begin{abstract} 
Student-generated metaphors about mathematics can provide insights into students’ attitudes, beliefs, identities, and experiences, but expert human assessment through thematic coding of these semantically complex metaphor responses is labor-intensive and difficult to scale. This study examines whether Low-Rank Adaptation (LoRA)-based supervised fine-tuning of Large Language Models (LLMs) can improve their performance on codebook-guided coding tasks for student mathematics metaphors. We utilized a human-coded corpus of 2,265 Grade 6–8 responses to food- and animal-based metaphor prompts and evaluated LLMs on two tasks: valence–intensity coding of students’ affective orientations toward mathematics and thematic coding of their metaphorical framings of mathematics. Two open-weight LLMs, DeepSeek-R1 1.5B and Mistral 7B, were evaluated before and after fine-tuning and compared with two proprietary LLMs, GPT-4o mini and GPT-5 mini. Results show that fine-tuning substantially improved the performance and run-to-run reliability of the open-weight LLMs across both tasks relative to their base versions, making the fine-tuned LLMs competitive with and often outperforming the proprietary LLMs. These findings suggest the potential of fine-tuned open-weight LLMs for scalable and automated AI-assisted measurement of students’ metaphor responses with competitive performance while maintaining local controllability and privacy-conscious deployment. 
\end{abstract} 

\section{Introduction}
Student-generated metaphors about mathematics offer insight into students’ mathematical beliefs, dispositions, and emotions by revealing how they interpret abstract ideas and experiences through familiar concrete domains \cite{cai2012mathematical,lakoff1980metaphors}. In mathematics education, these metaphors can capture both students’ affective orientations toward mathematics, such as whether they view it as enjoyable, frustrating, or threatening, and their broader framings of mathematics as useful, difficult, procedural, obligatory, or personally meaningful \cite{cai2012mathematical,schinck2008using}. For example, likening mathematics to a dog that acts as a companion and helps with everyday difficulties reflects a positive view of mathematics as useful and supportive, whereas describing mathematics as a mosquito that always comes back reflects a negative view of mathematics as annoying, unavoidable, and frustrating. These examples illustrate how metaphor responses can capture both affective valence and broader themes in students’ relationships with mathematics. Accurately identifying these affective and thematic dimensions can support more nuanced assessment of students’ mathematical dispositions, beliefs, and learning experiences and inform more responsive instruction \cite{deal2025accessing,schinck2008using, wegner2020metaphors}. However, such assessment typically relies on expert human coding, which is labor-intensive and requires specialized expertise and inferential judgment.  

Large Language Models (LLMs) have demonstrated strong natural-language understanding and inference capabilities, enabling them to apply predefined codebooks to open-ended responses and support scalable deductive qualitative coding \citep{liu2025qualitative, tai2024examination,xiao2023supporting}. LLM-based text classification and coding have been studied across political, social-science, and general Natural Language Processing (NLP) tasks \citep{alizadeh2024open,yu2023open}, and multi-code classification studies provide guidance for evaluating responses that require multiple codes \citep{chalkidis2020empirical,ma2025large}. These LLM capabilities make them promising for identifying and coding student metaphor responses, which requires inferring affective orientations and meaning-based codes from metaphor content \citep{deal2025accessing}. Two practical considerations motivate the present study’s approach to automated and scalable coding of student-generated metaphor responses. First, student responses may contain personally identifiable or contextually sensitive information, raising privacy concerns when such data are shared with third-party platforms to access proprietary LLMs \citep{pardo2014ethical,yan2024practical}. Second, although proprietary LLMs provide strong general-purpose language capabilities, they can limit researchers’ control over task-specific adaptation to specialized educational coding frameworks \citep{chung2024scaling,ouyang2022training}. Open-weight LLMs hosted on secure, controlled servers can be fine-tuned using human-assigned codes as supervision, enabling closer alignment with task-specific coding schemes while remaining competitive with proprietary models on specialized coding tasks \citep{alizadeh2025open,bucher2024fine,yu2023open,hu2022lora}. Given these considerations, the present study examines whether open-weight LLMs can accurately and reliably support codebook-guided coding of students’ mathematical metaphor responses across multiple layers of meaning. This study aims to address this gap by framing mathematics metaphor coding as a codebook-guided coding task and examining whether fine-tuning open-weight LLMs improves coding performance, where a commonly used parameter-efficient adaptation method, Low-Rank Adaptation (LoRA)-based supervised fine-tuning \citep{hu2022lora}, is utilized to align LLMs with human-coded metaphors. Two coding tasks are used to capture complementary dimensions: valence–intensity coding, a single-code task that codes the affective direction and strength of each response, and thematic coding, a multi-code task that identifies themes reflecting the underlying meanings students express about mathematics. The secure-server-deployable open-weight LLMs, DeepSeek-R1 1.5B \cite{deepseek_r1_distill_qwen_15b} and Mistral 7B \cite{mistral_7b_instruct_model_card}, were selected as representative lower-parameter models and evaluated both before and after supervised fine-tuning (SFT) \cite{ouyang2022training} for the metaphor-coding tasks. For comparison, two proprietary API-accessed OpenAI models, GPT-4o mini \cite{openai_gpt4o_mini}, a fast and cost-efficient general-purpose model, and GPT-5 mini \cite{openai_gpt5_mini}, a cost-efficient model with reasoning support for well-defined tasks, were evaluated under prompt-only conditions. 

This study advances AI-assisted mathematical measurement by demonstrating the potential of fine-tuned compact open-weight LLMs for scalable and privacy-conscious codebook-guided analysis of students’ mathematical metaphors, supporting more responsive mathematics instruction. 

\section{Related Work}

\textbf{Student Metaphors About Mathematics as Evidence of Attitudes and Beliefs.} Metaphors help learners make sense of mathematical ideas through familiar experiences \cite{LakoffNunez2000,lakoff1980metaphors}. In mathematics education, student metaphors have been used to examine affective dispositions, beliefs about mathematics, and experiences with school mathematics \cite{Yee2017,CaiMerlino2011,schinck2008using,Frid2001}. Root metaphors, such as mathematics as a structure, language, journey, or toolkit \cite{Noyes2006}, can reflect whether students view mathematics as useful, difficult, enjoyable, obligatory, disconnected, or conceptually meaningful. Analyzing both learning-oriented and self-referential metaphors about mathematics \cite{wegner2020metaphors} can provide useful information for both researchers and teachers to understand students' relationships with mathematics and inform instructional decisions that support the development of positive mathematical dispositions and identities. Because students' views of mathematics are intertwined with their learning experiences, their metaphors about mathematics and learning mathematics may overlap \cite{LatterellWilson2016} and can frequently encode multiple meanings. Thus, interpreting students' metaphors for mathematics can be a revealing if complex task, especially at scale, even for skilled researchers and teachers.  

\textbf{Codebook-Guided Qualitative Coding with LLMs.} Codebook-guided deductive qualitative coding uses predefined categories to identify meaningful patterns, a task in which LLM-based coding can serve as support for human interpretation \cite{hila2025assessing, tornberg2025large}. LLM-based approaches apply expert-developed codebooks to open-ended responses, but their effectiveness depends on the target construct, prompt design, and human validation procedures \cite{dunivin2025scaling,liu2025qualitative,tai2024examination,xiao2023supporting}. Methodologically, SFT improves the ability of LLMs to follow task-specific instructions and output formats \cite{chung2024scaling,ouyang2022training}, whereas LoRA enables efficient adaptation of open-weight LLMs through lightweight adapter modules rather than full parameter updates \cite{hu2022lora}. Empirical comparisons further show that fine-tuned smaller or open weight LLMs can be competitive with commercial prompt-only LLMs on constrained classification, annotation, and educational response-coding tasks \cite{kakarla2025comparing,alizadeh2025open,bucher2024fine,yu2023open}. Building on this work, our study examines whether compact open-weight LLMs can be fine-tuned for codebook-guided deductive coding of students' multi-layered mathematics metaphors. 

\section{Methods}
\textbf{Dataset.} Data were collected as part of a larger IRB-approved project (Protocol \#1730200-1) supporting teachers in teaching mathematics through problem posing, with informed consent/assent obtained from students and their parents/guardians. Grades 6--8 students completed two metaphor prompts: (a) ``If math were a food, it would be \underline{\hspace{1.2cm}} because \underline{\hspace{1.8cm}}.'' and (b) ``If math were an animal, it would be \underline{\hspace{1.2cm}} because \underline{\hspace{1.8cm}}.'' We used a human-coded corpus of 2,265 responses, including 1,139 food-based and 1,126 animal-based metaphors. As summarized in Table~\ref{tab:coding_scheme}, each response was coded through two complementary tasks: valence--intensity coding and thematic coding. Valence--intensity coding was a single-code task in which each response received one affective code from 1 (very negative) to 5 (very positive), with additional flags for incomplete or misinterpreted responses. Three human experts coded valence--intensity, with each response coded by two coders, yielding Krippendorff's $\alpha$ values of $0.85$, $0.84$, and $0.91$. Thematic coding was a qualitative multi-code task in which student responses could receive one or more fine-grained thematic codes when they expressed multiple underlying meanings (see Table~\ref{tab:coding_scheme}). The qualitative coding scheme was developed through thematic analysis of approximately 250 responses and refined through additional double coding and discussion; a further 100 responses reached 92\% exact-match agreement. These human-assigned codes served as references for training, fine-tuning, and evaluating the LLM-based coding models. More codebook details and coding examples are provided in  
\href{https://anonymous.4open.science/r/aime-con-2026-math-coding-36C9/AIME__LLM_for_Coding_Student_Metaphors__Appendix_Doc.pdf}{Appendix A}.

\begin{table}[ht!]
\centering
\tiny
\caption{Human coding scheme for valence--intensity and thematic coding.}
\label{tab:coding_scheme}
\renewcommand{\arraystretch}{1.05}
\setlength{\tabcolsep}{2.2pt}

\begin{tabularx}{\columnwidth}{@{}p{1.35cm}p{0.75cm}X@{}}
\toprule
\textbf{Task} & \textbf{Code} & \textbf{Rubric / Meaning} \\
\midrule

\multirow{7}{=}{\textbf{Valence--Intensity}}
& 1 & Very negative, e.g., disgusting; hate it \\
& 2 & Moderately negative, e.g., least favorite; dislike \\
& 3 & Neutral or ambivalent, e.g., okay; reader-dependent \\
& 4 & Moderately positive, e.g., like it \\
& 5 & Very positive, e.g., love it \\
& X & Cannot understand, incomplete, or blank; valence null \\
& M & Possible misunderstanding or literal response, used \emph{with} a numeric code 1--5 \\

\midrule

\multirow{26}{=}{\textbf{Thematic}}
& \multicolumn{2}{@{}l}{\textbf{A. Student's relationship to mathematics}} \\
& A11 & Familiar / comfortable / easy / accessible \\
& A12 & Interesting / cool / amazing / exciting / rare \\
& A13 & Enjoyable \\
& A14 & Beneficial to the individual \\
& A21 & Dangerous / threatening / scary \\
& A22 & Challenging / difficult / hard \\
& A23 & Unpleasant / disgusting / annoying / strange \\
& A24 & Boring \\
& A31 & Relationship changes over time or with experience \\
& A32 & Relationship depends on others' actions, teaching, or presentation \\
& A33 & Relationship varies by topic, task, or instance \\
& A34 & Includes positive and negative dimensions at the same time \\
& A35 & Some people like math and others do not \\
& A4  & Non-affective statement \\

\cmidrule(l){2-3}
& \multicolumn{2}{@{}l}{\textbf{B. Mathematics as a discipline}} \\
& B1 & Math is procedural \\
& B2 & Math requires precision / accuracy \\
& B3 & Math is about mathematical objects: numbers, shapes, equations, graphs \\
& B4 & Math is a collection of exercises / problems \\
& B5 & Math is for smart people \\

\cmidrule(l){2-3}
& \multicolumn{2}{@{}l}{\textbf{C. Mathematics and society}} \\
& C1 & Math is high status / important \\
& C2 & Math is useless / dumb \\
& C3 & Math is complex or unpredictable \\

\cmidrule(l){2-3}
& \multicolumn{2}{@{}l}{\textbf{D. Mathematics in school}} \\
& D1 & Math is a necessity / obligation \\
& D2 & Math requires persistence \\

\bottomrule
\end{tabularx}
\end{table} 

\textbf{LLMs Used.} All base LLMs used in this study are listed in Table~\ref{tab:llm_models}. We evaluated two proprietary LLMs, GPT-4o mini (cost-efficient model) and GPT-5 mini (reasoning-oriented model), and two locally deployable open-weight LLMs, DeepSeek-R1 1.5B and Mistral 7B. Both proprietary LLMs were accessed through an institutionally secured Microsoft Azure endpoint and evaluated as prompt-only baselines without task-specific fine-tuning. Both open-weight LLMs were deployed on a secure institutional server and evaluated under both base and fine-tuned conditions. 
\begin{table}[ht!]
\centering
\scriptsize
\caption{Base LLMs used in this study.}
\label{tab:llm_models}
\renewcommand{\arraystretch}{1.05}
\setlength{\tabcolsep}{1.8pt}
\begin{tabularx}{\columnwidth}{@{}Xlll@{}}
\toprule
\textbf{Model Name} & \textbf{Developer} & \textbf{Model Type} & \textbf{\# of Parameters} \\
\midrule
GPT-4o mini & OpenAI & Proprietary  & Not disclosed \\
GPT-5 mini & OpenAI & Proprietary & Not disclosed \\
DeepSeek-R1 1.5B & DeepSeek AI & Open-weight & 1.5B \\
Mistral 7B & Mistral AI & Open-weight & 7B \\
\bottomrule
\end{tabularx}
\parbox{\columnwidth}{\tiny\textit{Note.} 
``Proprietary'' refers to provider-hosted models accessed through an API without access to the underlying model weights. 
``Open-weight'' refers to models whose trained weights are available for local deployment and adaptation. 
Because OpenAI does not publicly disclose parameter counts for GPT-4o mini or GPT-5 mini, these models are reported by access type rather than exact model size.}
\end{table} 

\textbf{LLM Prompts for Coding.} 
All LLMs performed the coding task in a zero-shot, codebook-guided prompting setting. A system message defined the LLM-based agent as a codebook-guided metaphor-coding assistant, while the user message provided the prompt type, metaphor object, and student explanation. The LLM prompt also included an embedded codebook that constrained the allowable coding decisions. Specifically, the LLM-based assistant was asked to assign one valence--intensity code and one or more thematic codes for each response, following an explicit schema that included the valence code, valence category, thematic code(s), code-family(s), brief evidence, and rationale. The user-facing portion followed the general pattern: ``Use the codebook to code this student metaphor response. Prompt type: [Food/Animal]. Student metaphor object: [object]. Student explanation: [response]. Return only JSON using the required schema.'' The LLM prompt framed the coding task by strictly following the coding scheme shown in Table~\ref{tab:coding_scheme}. 

\textbf{Fine-tuning Setup for Open-Weight LLMs.} We fine-tuned DeepSeek-R1 1.5B and Mistral 7B using LoRA-based supervised fine-tuning, which kept the original model weights frozen while training small low-rank adapters to align model outputs with human-coded target labels \cite{hu2022lora}. Each training case paired a student's metaphor object and explanation with the corresponding human-assigned valence--intensity and root-metaphor semantic codes, enabling the models to learn expert coding decisions from human coded examples. Each LLM was fine-tuned using its native chat template. Specifically, the full prompt and student response were provided as input, while only prediction errors on the human-coded target output contributed to updating the LoRA adapters. Key training hyperparameters included a LoRA rank of 16, a scaling factor of 32, a dropout rate of 0.05, a maximum sequence length of 2,048 tokens, three training epochs, a per-device batch size of 8, two gradient accumulation steps, a learning rate of 0.0001, weight decay of 0.01, a warmup ratio of 0.03, cosine learning-rate scheduling, and a random seed of 42. Both LLMs were trained in 16-bit precision without 4-bit quantization, with additional details provided in \href{https://anonymous.4open.science/r/aime-con-2026-math-coding-36C9/AIME__LLM_for_Coding_Student_Metaphors__Appendix_Doc.pdf}{Appendix B}. 

\textbf{Data Splits and Evaluations.} The human-coded corpus comprised 2,265 student-metaphor responses and was split into training (783 food, 785 animal), validation (176 food, 174 animal), and held-out test sets (180 food, 167 animal), with approximately balanced food- and animal-based responses across subsets. All LLMs were evaluated on the two coding tasks using the same held-out test set, coding prompt structure, valid-output checking rules, and evaluation metrics. The temperature of the LLMs was set to 0 to reduce random variations. For GPT-5-mini, explicit temperature control was not supported by the provider API \cite{openai_temperature,openai_gpt5_compatibility}. Human-assigned codes served as the gold-standard reference. For the single-code valence--intensity task, we used: 1) \emph{Accuracy}, which measures the proportion of responses for which the predicted valence code exactly matches the human-assigned code; 2) \emph{Macro-F1}, which calculates \textit{F1} for each possible valence code and averages them equally, so each valence code contributes equally to the final score \cite{takahashi2022confidence}; and 3) \emph{Quadratic Weighted Kappa} (QWK), which measures ordinal agreement between predicted and human-assigned valence codes while assigning larger penalties to larger code differences (this is appropriate because the valence--intensity coding task uses an ordered 1--5 scale ranging from very negative to very positive) \cite{cohen1968weighted}. For the multi-coding situation \cite{zhang2013review,read2010scalable,yang1999evaluation} in thematic coding, we used: 1) \emph{Subset Accuracy}, which measures the proportion of responses for which the full set of predicted codes exactly matches the human-assigned codes \cite{zhang2013review}; 2) \emph{Macro-F1}, which calculates \textit{F1} for each code family and averages them, giving equal weight to both frequent and infrequent codes \cite{takahashi2022confidence,read2010scalable}; and 
3) \emph{Micro-F1}, which pools true positives, false positives, and false negatives across all code families before calculating \textit{F1}, giving more influence to frequent codes \cite{takahashi2022confidence,read2010scalable}. Each LLM was evaluated three times under identical conditions, with results averaged across runs. 

\section{Results} 

\begin{figure*}[ht!]
    \centering
    \includegraphics[width=\linewidth]{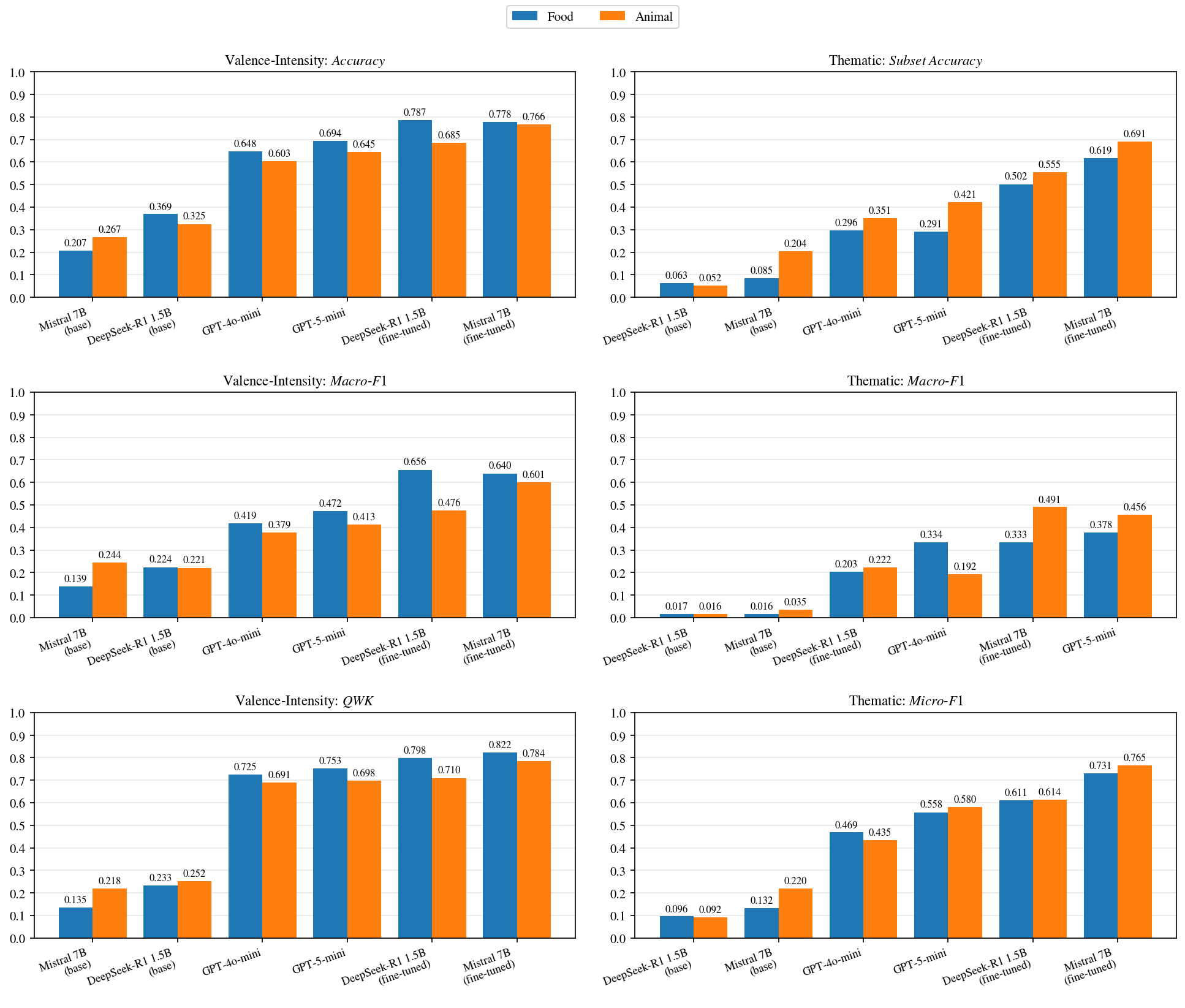}
    \caption{Coding performance across LLMs for food- and animal-based metaphor responses.}
    \label{fig:semantic_performance}
\end{figure*}

\textbf{Coding Performance Across LLMs.} 
Figure~\ref{fig:semantic_performance} presents the coding performance of
all LLMs on valence--intensity coding and thematic coding
for both food- and animal-based metaphor responses. Overall, fine-tuning
substantially improved the performance of the two open-weight LLMs,
DeepSeek-R1 1.5B and Mistral 7B, over their corresponding base versions across
both response subsets and both coding tasks. For valence--intensity coding,
fine-tuned DeepSeek-R1 1.5B improved in \textit{Accuracy} from $0.369$ to
$0.787$ on the food subset and from $0.325$ to $0.685$ on the animal subset,
while fine-tuned Mistral 7B improved from $0.207$ to $0.778$ on the food subset
and from $0.267$ to $0.766$ on the animal subset. Similar gains were observed
for thematic coding. For example, fine-tuned DeepSeek-R1 1.5B
improved in \textit{Subset Accuracy} from $0.063$ to $0.502$ on the food subset
and from $0.052$ to $0.555$ on the animal subset. The \textit{F1}-based and agreement metrics also showed the coding performance gains. For example, fine-tuned Mistral 7B improved in
valence--intensity \textit{Macro-F1} from $0.139$ to $0.640$ on the food subset
and from $0.244$ to $0.601$ on the animal subset, while its valence--intensity
\textit{QWK} increased from $0.135$ to $0.822$ on the food subset and from
$0.218$ to $0.784$ on the animal subset. For thematic coding,
fine-tuned Mistral 7B improved in \textit{Micro-F1} from $0.132$ to $0.731$ on
the food subset and from $0.220$ to $0.765$ on the animal subset, while
fine-tuned DeepSeek-R1 1.5B improved from $0.096$ to $0.611$ on the food subset
and from $0.092$ to $0.614$ on the animal subset. Compared with the proprietary prompt-only baselines, the fine-tuned open-weight
LLM models achieved competitive or stronger performance across most metrics. In
particular, fine-tuned Mistral 7B exceeded both GPT-4o-mini and GPT-5-mini on
every reported metric for both the food and animal subsets, with the sole exception of
thematic \textit{Macro-F1} on the food subset, where it was essentially tied with
GPT-4o-mini ($0.333$ vs.\ $0.334$) and below GPT-5-mini ($0.378$). Fine-tuned
DeepSeek-R1 1.5B also outperformed the proprietary baselines on most metrics,
with the main limitation appearing in thematic \textit{Macro-F1}:
It reached $0.203$ on the food subset, below both GPT-4o-mini ($0.334$) and
GPT-5-mini ($0.378$), and $0.222$ on the animal subset, below GPT-5-mini
($0.456$) but above GPT-4o-mini ($0.192$). This pattern suggests that fine-tuning improved
overall task performance, while performance on less frequent thematic
codes remained more uneven. Detailed significance test results for the coding
performance gains are provided in \href{https://anonymous.4open.science/r/aime-con-2026-math-coding-36C9/AIME__LLM_for_Coding_Student_Metaphors__Appendix_Doc.pdf}{Appendix C}. 

\textbf{Run-to-Run Reliability.} To assess the stability of LLM outputs across repeated runs, we computed Krippendorff's $\alpha$ \cite{krippendorff2011computing} for valence--intensity coding and mean pairwise exact-match agreement for thematic coding, with results shown in Table~\ref{tab:run_reliability_alpha}. For valence--intensity coding, most LLMs showed high run-to-run stability above 0.90, except the base DeepSeek-R1 1.5B model, which was less stable ($\alpha = 0.592$). For thematic coding, fine-tuning likewise improved run-to-run reliability: DeepSeek-R1 1.5B increased from $57.6\%$ to $96.0\%$ exact-code agreement, and Mistral 7B increased from $98.1\%$ to $99.4\%$. Both fine-tuned open-weight models also outperformed the commercial prompt-only models, including GPT-4o-mini ($87.2\%$) and GPT-5-mini ($64.4\%$). These results indicate that fine-tuning substantially improves the run-to-run stability of open-weight LLM coding in both coding tasks. 

\begin{table}[ht!]
\tiny
\centering
\caption{Run-to-run reliability across the three independent runs. Valence--intensity coding is measured with
Krippendorff's $\alpha$ (ordinal), and thematic coding is measured with mean pairwise exact-match agreement
across runs at the exact root-code level.}
\label{tab:run_reliability_alpha}
\renewcommand{\arraystretch}{1.15}
\setlength{\tabcolsep}{3pt}

\begin{tabularx}{\columnwidth}{
@{}
>{\raggedright\arraybackslash}p{2.85cm}
>{\centering\arraybackslash}p{2.50cm}
>{\centering\arraybackslash}p{1.85cm}
@{}}
\toprule
\textbf{Model} & \textbf{Valence--Intensity} & \textbf{Thematic} \\
 & ($\alpha$) & (exact match) \\
\midrule
GPT-4o-mini                   & 0.993 & 87.2\% \\
GPT-5-mini                    & 0.949 & 64.4\% \\
DeepSeek-R1 1.5B (base)       & 0.592 & 57.6\% \\
DeepSeek-R1 1.5B (fine-tuned) & 0.992 & 96.0\% \\
Mistral 7B (base)             & 0.948 & 98.1\%$^{\dagger}$ \\
Mistral 7B (fine-tuned)       & 1.000 & 99.4\% \\
\bottomrule
\end{tabularx}

\par\vspace{3pt}
\parbox{\columnwidth}{\tiny\textit{Note.} The three runs are treated as raters; thematic agreement requires the exact root-code set to
match between two runs. $^{\dagger}$This model emitted the single root code $\{$A22$\}$ for roughly $83\%$ of
responses (all within family A) across all three runs, so its high agreement reflects a near-constant output
rather than genuine discrimination across items.}
\end{table}

\textbf{Coding Performance Gains Across Codes.}
\label{sec:label_based_gains}
To examine where fine-tuning helps, Table~\ref{tab:rq2_perclass_acc_f1} reports code-level
base-to-fine-tuned changes for the two open-weight LLMs, with per-level results for
valence--intensity coding and per-family (grouped exact-code) results for thematic coding. The
overall pattern is consistent: Fine-tuning improved nearly every code, with the largest gains
concentrated where the base models were weakest. For valence--intensity, per-level \textit{F1}
rose on every code for both models. DeepSeek-R1~1.5B gained most on the very-positive code~``5'', where \textit{F1} rose from $0.067$ to $0.667$ ($+0.600$) and \textit{Accuracy}
from $0.083$ to $0.750$ ($+0.667$), while Mistral~7B improved most on the neutral code~``3'',
where \textit{F1} rose from $0.178$ to $0.793$ ($+0.614$) and \textit{Accuracy} from $0.105$ to
$0.823$ ($+0.718$). The only decline was Mistral's Code ``1'' \textit{Accuracy}, which fell from
$0.962$ to $0.654$ because the base model indiscriminately over-predicted the negative extreme. Its Code ``1'' \textit{F1} still rose
sharply, from $0.210$ to $0.723$ ($+0.514$), once precision improved. For thematic coding, the
largest gain lay within the dominant family~A (relationship to math): The base LLMs scored very
low at the exact-code level (grouped exact-code \textit{F1} of $0.066$ for DeepSeek-R1~1.5B and
$0.119$ for Mistral~7B), and fine-tuning raised this to $0.508$ and $0.744$ ($+0.442$ and
$+0.625$), the largest thematic coding performance gain for each model. The rarer discipline-, society-, and
school-level families~B, C, and~D were essentially absent from the base outputs
($\textit{F1}=0.000$ throughout) and improved more modestly (B: $0.252/0.441$; C: $0.233/0.536$;
D: $0.000/0.333$), with DeepSeek-R1~1.5B still unable to recover family~D ($0.000$). Overall,
fine-tuning's benefit comes chiefly from the dominant family~A, where it lifts exact-code
performance from near-zero to $0.508$--$0.744$, while on the rare families~B, C, and~D it enables
the models to produce codes the base versions had essentially never assigned.  

\renewcommand{\up}[1]{\ensuremath{+#1}}
\renewcommand{\dn}[1]{\ensuremath{-#1}}
\renewcommand{\upb}[1]{\ensuremath{\mathbf{+#1}}}
% also needs: \usepackage{booktabs,multirow,graphicx}

\begin{table*}[ht!]
\tiny
\centering
\caption{Base to fine-tuned per-class results on the held-out test set ($n=347$, 3-run mean, food and animal responses combined). 
For valence-intensity levels we report per-level \textit{Accuracy} (= recall on that level) and per-level \textit{F1};
for thematic coding families we report the \emph{grouped exact-code} \textit{F1} (each family value is the
equal-weight mean of the exact root-code \textit{F1}s within that family, scored by exact-code match).
$\Delta$ is fine-tuned minus base; the largest gain per model is in bold.}
\label{tab:rq2_perclass_acc_f1}
\renewcommand{\arraystretch}{1.12}
\setlength{\tabcolsep}{4.2pt}

\resizebox{\textwidth}{!}{%
\begin{tabular}{cccccccccc}
\toprule
\multirow{2}{*}{\textbf{Coding Task}}
& \multirow{2}{*}{\textbf{Code}}
& \multirow{2}{*}{\textbf{$n$}}
& \multirow{2}{*}{\textbf{Metric}}
& \multicolumn{3}{c}{\textbf{DeepSeek-R1 1.5B}}
& \multicolumn{3}{c}{\textbf{Mistral 7B}} \\
\cmidrule(lr){5-7}\cmidrule(l){8-10}
& & & & \textbf{Base} & \textbf{Fine-tuned} & \textbf{$\Delta$} & \textbf{Base} & \textbf{Fine-tuned} & \textbf{$\Delta$} \\
\midrule

\multirow{10}{*}{\textbf{Valence--Intensity}}
 & \multirow{2}{*}{1} & \multirow{2}{*}{26} & \textit{Accuracy}
 & 0.282 & 0.436 & \up{0.154}
 & 0.962 & 0.654 & \dn{0.308} \\
 & & & \textit{F1}\textsuperscript{a}
 & 0.307 & 0.553 & \up{0.246}
 & 0.210 & 0.723 & \up{0.514} \\

\cmidrule(l){2-10}
 & \multirow{2}{*}{2} & \multirow{2}{*}{120} & \textit{Accuracy}
 & 0.181 & 0.753 & \up{0.572}
 & 0.272 & 0.767 & \up{0.494} \\
 & & & \textit{F1}\textsuperscript{a}
 & 0.257 & 0.754 & \up{0.497}
 & 0.302 & 0.793 & \up{0.491} \\

\cmidrule(l){2-10}
 & \multirow{2}{*}{3} & \multirow{2}{*}{130} & \textit{Accuracy}
 & 0.503 & 0.836 & \up{0.333}
 & 0.105 & 0.823 & \upb{0.718} \\
 & & & \textit{F1}\textsuperscript{a}
 & 0.454 & 0.786 & \up{0.331}
 & 0.178 & 0.793 & \upb{0.614} \\

\cmidrule(l){2-10}
 & \multirow{2}{*}{4} & \multirow{2}{*}{61} & \textit{Accuracy}
 & 0.421 & 0.650 & \up{0.230}
 & 0.153 & 0.754 & \up{0.601} \\
 & & & \textit{F1}\textsuperscript{a}
 & 0.344 & 0.680 & \up{0.336}
 & 0.254 & 0.730 & \up{0.476} \\

\cmidrule(l){2-10}
 & \multirow{2}{*}{5} & \multirow{2}{*}{8} & \textit{Accuracy}
 & 0.083 & 0.750 & \upb{0.667}
 & 0.167 & 0.750 & \up{0.583} \\
 & & & \textit{F1}\textsuperscript{a}
 & 0.067 & 0.667 & \upb{0.600}
 & 0.255 & 0.706 & \up{0.451} \\

\midrule

\multirow{4}{*}{\textbf{Thematic Family}}
 & A & 285 & \textit{F1}\textsuperscript{b}
 & 0.066 & 0.508 & \upb{0.442}
 & 0.119 & 0.744 & \upb{0.625} \\
 & B & 45 & \textit{F1}\textsuperscript{b}
 & 0.000 & 0.252 & \up{0.252}
 & 0.000 & 0.441 & \up{0.441} \\
 & C & 22 & \textit{F1}\textsuperscript{b}
 & 0.000 & 0.233 & \up{0.233}
 & 0.000 & 0.536 & \up{0.536} \\
 & D & 6 & \textit{F1}\textsuperscript{b}
 & 0.000 & 0.000 & \up{0.000}
 & 0.000 & 0.333 & \up{0.333} \\

\bottomrule
\end{tabular}%
}

\begin{flushleft}
\scriptsize
\textit{Note.} \textsuperscript{a}For valence levels, the reported \textit{F1} is the per-level (macro-scheme)
\textit{F1}: the equal-weight average of the five per-level \textit{F1} values equals the task-level valence
\emph{Macro-F1}. Per-level Accuracy equals recall on that level (the share of responses humans coded at that
level that the model also coded there). Two responses labeled X were excluded from the valence--intensity
analysis because they did not have valid numeric 1--5 codes.
\textsuperscript{b}For thematic families, the reported \textit{F1} is the \emph{grouped exact-code} \textit{F1}:
Predictions are scored by exact root-code match (e.g., A11 vs.\ A12 is a miss), and each family value is the
equal-weight mean of the exact-code \textit{F1}s for the root codes in that family (A:~13, B:~5, C:~2, D:~2
codes; $n$ is the number of test responses carrying that family). Because families differ in size and each is
weighted equally, these family means do \emph{not} average to the flat exact-code thematic \emph{Macro-F1}.
Three task-level metrics are not shown here because they cannot be decomposed to a single class: \emph{QWK}
requires the full ordinal 1--5 valence scale; \emph{Subset Accuracy} is defined over the entire predicted
root-code set; and \emph{Micro-F1} pools decisions across all codes.
\end{flushleft}
\end{table*}  

\section{Discussion}
These findings suggest that the gains in coding performance of the fine-tuned open-weight LLMs reflect improved task-specific alignment with expert human-annotated codes for the codebook-guided valence--intensity and thematic coding tasks. By training on human-coded metaphor responses, the LLMs learned from human expert examples of how student responses should be mapped to valence--intensity codes and thematic codes. This SFT enabled the open-weight LLMs to more closely reproduce expert coding decisions and human-assigned target codes. This task-specific adaptation may help explain why fine-tuned DeepSeek-R1 1.5B and Mistral 7B became competitive with or outperformed the proprietary models, including GPT-4o mini and GPT-5 mini, on these tasks. This finding also aligns with related work showing that task-specific fine-tuning, including LoRA-based adaptation, can help smaller open-weight models trained on human-coded datasets become competitive with proprietary LLMs on specialized classification, annotation, and educational response-coding tasks \citep{alizadeh2024open,bucher2024fine,kakarla2025comparing,hu2022lora,yu2023open}. Additionally, the run-to-run reliability results indicate that the fine-tuned LLMs produced stable outputs across repeated inference runs. In mathematical measurement, such stability supports more consistent coding of the same student responses, reduces random measurement variation, and improves the reproducibility of LLM-based coding. 

The code-level fine-tuning results show that gains were uneven across valence--intensity levels and thematic families. This suggests that fine-tuning did not simply improve LLM-powered coding performance uniformly but instead helped the LLMs learn some parts of the codebook more effectively than others. One plausible task-level explanation is that different coding tasks require different forms of inference: Valence--intensity coding depends on affective inference about the polarity and strength of students’ attitudes, whereas thematic coding requires conceptual and semantic inference to map a response onto one or more code categories. These code-level differences may reflect both the models’ task-specific capabilities developed during pretraining and the fine-tuning process, through which models may learn some expert-defined code boundaries more effectively than others depending on how well they connect linguistic cues in individual responses to each coding category. This observation of code-level variation is also reflected in other work showing that LLM coding performance can differ across target constructs, coding contexts, and multi-code settings \citep{liu2025qualitative,ma2025large}. Further investigation is still needed to better understand the sources of coding difficulty and variation across individual codes and tasks. 

% Though some differences between the food and animal objects appeared (e.g., slightly higher valence accuracy on food and slightly higher family agreement on animal), they were small and inconsistent across metrics, and no clearly significant subgroup effect could be observed (see \href{https://anonymous.4open.science/r/aime-con-2026-math-coding-36C9/AIME__LLM_for_Coding_Student_Metaphors__Appendix_Doc.pdf}{Appendix B} and the significance tests in \href{https://anonymous.4open.science/r/aime-con-2026-math-coding-36C9/AIME__LLM_for_Coding_Student_Metaphors__Appendix_Doc.pdf}{Appendix C}). These findings suggest that label-level analysis is necessary for understanding where fine-tuned LLMs align well with expert coding decisions. 

% \textbf{Impacts Paragraph.} 
These findings suggest that fine-tuned LLMs can support scalable mathematical measurement of open-ended student responses while maintaining alignment with human-developed mathematics education codebooks. Open-weight LLMs deployed on secure institutional servers offer a privacy-conscious alternative to proprietary systems and can be adapted to specialized coding schemes using human-assigned codes. The code-level results further provide diagnostic value by identifying affective and thematic categories that are more or less reliably captured by LLM coding. Thus, fine-tuned LLMs can serve as codebook-aligned research assistants for scalable qualitative analysis in mathematics education. 

\textbf{Limitations and Future Work.}
We note four main limitations. First, all LLMs were evaluated only with temperature $=0$, so future work should examine other decoding settings \citep{agarwal2024understanding,zhang2024edt}. Second, the corpus was limited to Grades 6--8 food- and animal-based mathematics metaphors, limiting generalizability. Third, rare thematic families remained less stable, suggesting the need for targeted data collection or class-balanced fine-tuning. Fourth, this study did not fully explain why performance differed across code levels, which should be further investigated in future work. 

\section{Conclusion}
This study examined whether LoRA-based supervised fine-tuning can improve LLM-based coding of student mathematics metaphors, statements with multiple levels of meaning that can offer insight into students' mathematical affect and beliefs about mathematics. Using a human-coded corpus of Grade 6--8 food- and animal-based metaphor responses, we evaluated two open-weight LLMs, DeepSeek-R1 1.5B and Mistral 7B, and two proprietary LLMs, GPT-4o mini and GPT-5 mini, on two coding tasks: valence--intensity coding and thematic coding. Results showed that fine-tuning on open-weight LLMs substantially improved both the coding performance and run-to-run stability of the open-weight models, with fine-tuned Mistral 7B achieving the strongest overall performance and fine-tuned DeepSeek-R1 1.5B showing large gains over its base version. Per-code analysis further showed that fine-tuning improved performance across many valence--intensity and thematic categories, although gains varied by code. These findings suggest that compact open-weight LLMs, when trained on human-coded examples, can provide a scalable and locally deployable approach for supporting codebook-guided deductive coding of thematically and semantically complex student responses such as mathematical metaphors in mathematics education. This work points to the potential of fine-tuned LLMs to support AI-assisted measurement of students' mathematical attitudes, beliefs, and experiences from open-ended responses. 

\section*{Acknowledgments}
This work was supported by the National Science Foundation under Grant DRL-2101552. The views expressed are those of the authors and do not necessarily reflect those of the Foundation. We also thank Prof. Ning Wang of Texas A\&M University for her valuable feedback. 

\bibliography{custom}

% \appendix

% \section{Example Appendix}
% \label{sec:appendix}

\newcolumntype{Y}{>{\centering\arraybackslash}X}
\newcolumntype{L}[1]{>{\raggedright\arraybackslash}p{#1}}

\end{document}